\documentclass[aps,prb,preprint,showpacs,showkeys,superscriptaddress]{revtex4}
\pdfoutput=1

\usepackage{amsmath}
\usepackage{amssymb}
\usepackage{graphicx}

\begin{document}

\title{Determining the energetics of vicinal perovskite oxide surfaces}

 \author{Werner A. \surname{Wessels}}
 \affiliation{Faculty of Science and Technology and MESA$^+$ Institute for 
Nanotechnology, 
Inorganic Materials Science, University of Twente, P.O. Box 217, 7500 AE, 
Enschede, The Netherlands.}

 \author{Tjeerd R.J. \surname{Bollmann}}
 \affiliation{Faculty of Science and Technology and MESA$^+$ Institute for 
Nanotechnology, 
Inorganic Materials Science, University of Twente, P.O. Box 217, 7500 AE, 
Enschede, The Netherlands.}

 \author{Gertjan \surname{Koster}}
 \affiliation{Faculty of Science and Technology and MESA$^+$ Institute for 
Nanotechnology, 
Inorganic Materials Science, University of Twente, P.O. Box 217, 7500 AE, 
Enschede, The Netherlands.}

\author{Harold J.W. \surname{Zandvliet}}
 \affiliation{Faculty of Science and Technology and MESA$^+$ Institute for 
Nanotechnology, 
Physics of Interfaces and Nanomaterials, University of Twente, P.O. Box 217, 
7500 AE, Enschede, The Netherlands.}

\author{Guus \surname{Rijnders}}
 \affiliation{Faculty of Science and Technology and MESA$^+$ Institute for 
Nanotechnology, 
Inorganic Materials Science, University of Twente, P.O. Box 217, 7500 AE, 
Enschede, The Netherlands.}
 

\begin{abstract}
The energetics of vicinal SrTiO$_3$(001) and DyScO$_3$(110), prototypical 
perovskite vicinal surfaces, has been studied using topographic atomic force 
microscopy imaging. The kink formation and strain relaxation energies are 
extracted from a statistical analysis of the step meandering. Both perovskite 
surfaces have very similar kink formation energies and exhibit a similar 
triangular step undulation. Our experiments suggest that the energetics of 
perovskite oxide surfaces is mainly governed by the local oxygen coordination.
\end{abstract}

\maketitle

\section{Introduction}

The perovskite oxides are a fascinating class of material, due to their 
wealth in 
available physical properties, such as superconductivity, ferromagnetism, 
ferro- and dielectricity. Despite their abundance of functional properties, the 
ABO$_3$ perovskite platform shares a common crystal structure with similar 
lattice parameters. For application within oxide thin film devices, grown 
multilayer (perovskite) heterostructures typically contain at least one 
functional active layer which is then directly supported onto a substrate, or 
with a bottom electrode layer in between film and substrate.
In these (ultra)thin film structures enormous strains might be beneficial, as 
they can result in unanticipated functional properties such as altering $T_c$ 
of ferromagnetic and superconducting materials \cite{Beach1993, 
Sato1997, Gan1998, Bozovic2002}. However, 
uncontrolled strain relaxation might also result in destructive cracking 
\cite{Beuth1992, Xia2000, Morito2005} or threading dislocation cascades 
\cite{Ernst1998, Oh2004, Wang2008} within thin films.
It is therefore of utmost importance to identify and quantify strain relaxation 
behavior within (ultra)thin films. As we will demonstrate furtheron, strain 
relaxation phenomena are easily identified at the thin film interface.
Besides strain relaxation, the kink formation energy is an important 
parameter in thin film growth, 
facilitating nucleation during growth and thereby determining the resulting 
growth to be either of rough 3D or 2D stepflow character.
The formation of steps usually introduces surface stress, influencing the step 
edge morphology. In view of its technological relevance, most of the 
step-related surface studies have been focused so far on silicon as a model 
system \cite{zandvliet2000,bartelt1992}.

In this paper, we describe a method to determine both, the strain 
relaxation energy together with the step edge formation energy by the use of 
Atomic Force Microscopy (AFM) as it can image the surface topography 
irrespective of its band gap. 
To our knowledge, this study is unique in the field of complex oxides as it is 
a quantitative study to extract energetic parameters using AFM where AFM 
studies reported in literature so far 
\cite{nishimura1999,connell2012,sanchez2014} have been of qualitative 
character. 
By imaging the meandering of step edges, which are determined by the 
step free energy, the kink formation energy can be determined 
\cite{zandvliet1995}.
As in mono-metal oxides the geometric structure at the surface is 
considered to be a continuation of the atomic arrangement of building blocks 
\cite{victor1996}, we assume in our analysis that for the more complex 
perovskite oxides the ABO$_3$ building block is the elementary unit to describe 
the surface energetics.

This paper is arranged as follows. First, we describe the experimental details 
with emphasis on the surface preparation to ensure the surface under study is 
in thermal equilibrium. Next, we explain the use of correlation functions to 
extract the kink energy ($E_{kink}$) and strain relaxation energy constant 
($C$) from the observed step meandering in topographical AFM images. Finally, 
we discuss the use of this procedure on two prototypical perovskite surfaces, 
i.e. SrTiO$_3$(001) and DyScO$_3$(110). The selected two systems have a lattice 
mismatch of 1.0\% at room temperature, however, are known to result upon 
epitaxial growth of the first onto the latter in films with 
a high degree of uniformity and structural perfection in comparison to typical 
films \cite{Haeni2004} indicative of their ability to accomodate strain well. 
We conclude by summarizing the observed foundings and argue how the reported 
implications are anticipated to be of 
generic character.

\section{Experimental details}
To quantify the strain relaxation energy, we image the step meandering of 
vicinal perovskite surfaces on standard commercial available 5$\times$5 
mm$^{2}$ samples of SrTiO$_{3}$(001) and DyScO$_{3}$(110) 
using topographical AFM images.
SrTiO$_{3}$(001) and DyScO$_{3}$(110) samples are often used substrates in thin 
film growth studies, as they are chemically stable and are known to have a low 
defect density. In addition, these materials have a well-defined surface of 
alternating AO and BO$_{2}$ planes, which is neutral in the case of 
SrTiO$_{3}$(001) and polar for DyScO$_{3}$(110). 
Furthermore, SrTiO$_{3}$(001) has a pseudo cubic lattice constant of 
$a_0$=3.905\AA\, whereas DyScO$_{3}$(110) has a nearly square surface mesh with 
an inplane lattice spacing of $a_0$=3.95\AA\ in the perovskite oxide 
spectrum \cite{schlom2008}. 
The assumption of a cubic unit cell for both was used throughout our 
analysis. 

Prior to annealing, all samples were ultrasonically rinsed in 
acetone and ethanol both for 10 minutes. 
In order to achieve single terminated BO$_2$ 
surfaces, we applied chemical etching procedures as described in detail 
elsewhere \cite{koster1998,kleibeuker2010}. The samples were then annealed in a 
tube furnace using an O$_{2}$ flow of 150~l/h at a temperature of 1273~K. The 
AFM topographic images 
were taken in ex-situ tapping 
mode (TM). The fast scan direction was aligned perpendicular to the surface 
steps in order to optimize image analysis and to prevent tip artifacts in the 
images. 

\begin{figure}
\centering\includegraphics[width=0.7\linewidth]{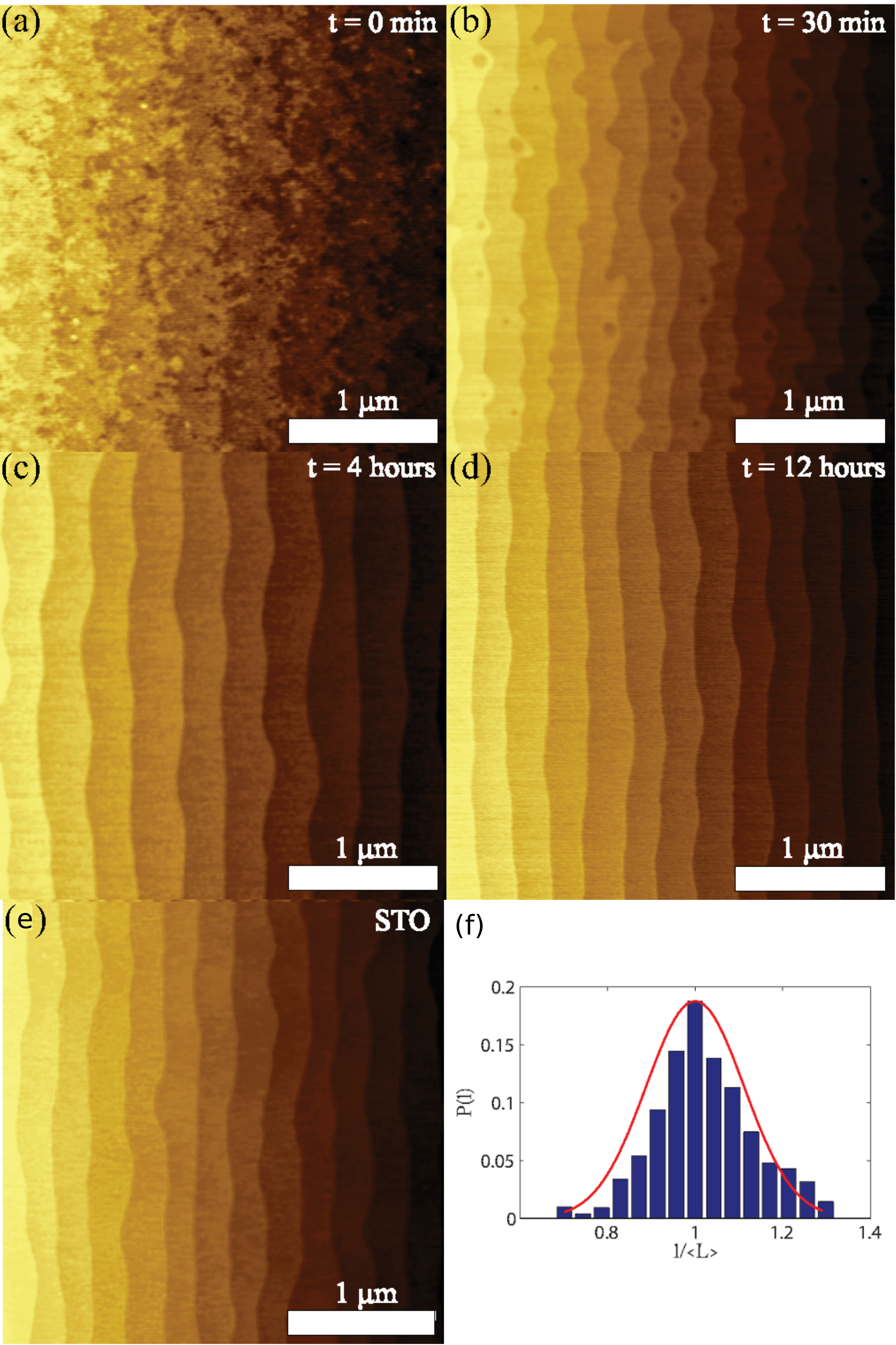}
\caption{\emph{Ex-situ} TM-AFM images of DyScO$_{3}$(110) surface at room 
temperature of a) an as-received surface, 
b) surface after 30 min annealing at 1273 K, c) surface after 4 hours annealing 
at 
1273 K and d) surface after 12 hours annealing at 1273 K, respectively. Note 
that 
these samples were cut from the same crystal.
e) \emph{Ex-situ} TM-AFM image of SrTiO$_3$(001) surface at thermodynamic 
equilibrium. $<L>$=250~nm.
f) Normalized terrace width distribution P($l$) vs. terrace width $l$ of 
the DyScO$_{3}$(110) surface annealed for 4~hours at 1273~K. An average terrace 
width $<L>$ = 760~a$_{0}$ is found. The narrow width $\sigma$ = 86 $a_{0}$ and 
asymmetry are indicative for entropic and energetic step-step interactions.}
\label{fig:1}
\end{figure} 
For the statistical analysis described below, it is of utmost importance that 
the surface is prepared at thermal equilibrium. By varying the annealing 
time and temperature followed by AFM imaging, we study the evolution of the 
step-edges to find the conditions for which the surface is at equilibrium. 
Fig.~\ref{fig:1}(a) shows the polished as-received surface, exhibiting 
disordered step edges, without a well-defined step distribution. Upon 30 
minutes of annealing as described above, the steps become 
visible, but still reveal vacancy islands in the terraces as well as 
protrusions along the step edges, indicative of its non-equilibrium state, see 
Fig.~\ref{fig:1}(b). 
Prolonged annealing up to 4~hours results in meandering of the surface steps 
and a 
well-defined and narrow terrace width distribution, see Fig.~\ref{fig:1}(c) and 
(f). No vacancy islands are present in the terraces.
Typically, the average terrace width ($<L>$) and its standard 
distribution ($\sigma$) are measured to study the strength of the step-step 
interactions.
To verify the terrace width, its distribution P($l$) versus the terrace width 
$l$ is plotted in Fig.~\ref{fig:1}(f). 
The data can nicely be described by the fitted Gaussian, neglecting the 
weak shoulders and the slight asymmetry in the distribution. An average terrace 
width $<L>$ of 760~a$_{0}$ (corresponding to 300~nm) is found for the 
DyScO$_3$(110) surface in Fig.~\ref{fig:1}(c). The standard distribution 
($\sigma$) of this 
distribution, only 86~$a_{0}$ (corresponding to 34~nm), is relatively small 
and indicative for step-step interactions. This very narrow distribution of 
DyScO$_{3}$(110) compared to silicon can be explained by the coherent behavior 
of the DyScO$_{3}$(110) step-edges \cite{zandvliet2000}. The shoulder on the 
right of the average in the distribution is most probably due to step-step 
repulsion.

Continuing annealing up to 12~hours, see Fig.~\ref{fig:1}(d), reveals no 
significant 
change of the surface or step edge correlation function analysis as described 
below, a clear signature of a surface at thermal equilibrium.
The same procedure was used to prepare and verify the SrTiO$_3$(001) surface at 
thermal equilibrium, see Fig.~\ref{fig:1}(e).
Note that the length scale at which thermal equilibrium is reached scales as 
$t^{\alpha}$, where $t$ is the time and $\alpha$ an exponent that depends on the 
exact mass transport mechanism \cite{mullins1957,mullins1959,mullins1963}. An 
exponent $\alpha$ = $\frac{1}{2}$ describes mass exchange between the step edge 
and adatoms on the terrace, whereas $\alpha$ = $\frac{1}{4}$ describes mass 
transport along step edges \cite{pimpinelli1993}. The large correlation length 
determined from the experimental obtained data ($\lambda\approx$~2100~$a_{0}$, 
see below) suggests that a non-equilibrium step instability is very unlikely 
\cite{wu1993,bales1990}.

\section{Statistical analysis}
\begin{figure}
\centering\includegraphics[width=0.8\linewidth]{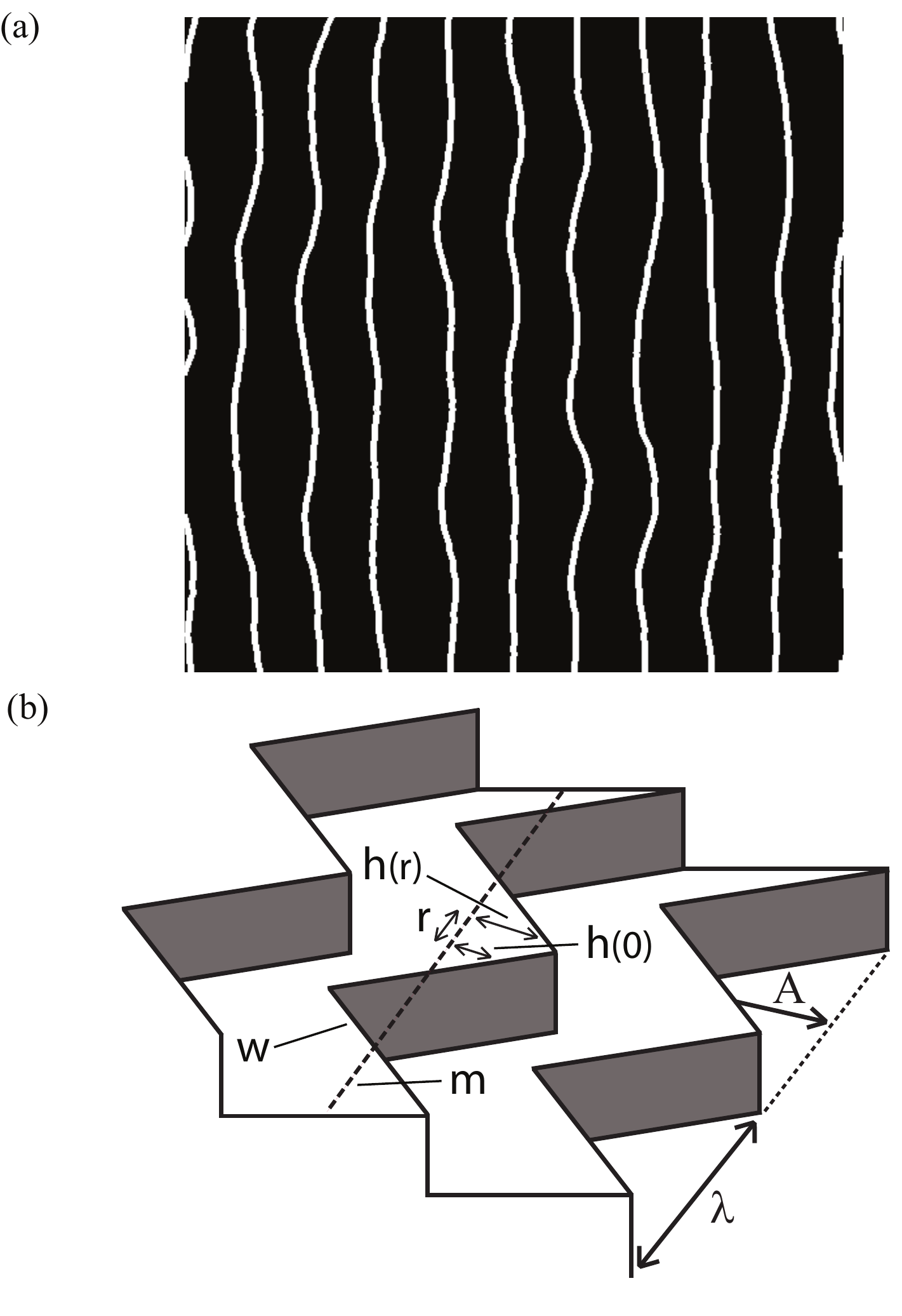}
\caption{Statistical analysis on a 512$\times$512 pixels$^{2}$ \emph{ex-situ} 
TM-AFM image. 
a) The Canny method applied to an AFM image of DyScO$_{3}$(110) to obtain the 
step edge profiles (displayed here in white).
b) Schematic representation of a triangular step edge undulation. The wavy $w$ 
line represents a step edge of the surface whereas its mean line is indicated by 
$m$.}
\label{fig:2}
\end{figure}

In order to extract the kink energy $E_{kink}$ and strain relaxation energy 
from the meandering step edges in the topographical AFM images, step edge 
analysis was performed on the AFM images by first applying a Gaussian filter to 
reduce noise. Subsequently, leveling of the height data was done by plane 
fitting the topography. Step edges were then detected by applying the Canny 
filter \cite{canny1986}, a multi-stage algorithm to detect a wide range of 
edges. Fig.~\ref{fig:2}(a) represents a processed AFM topographical image of a 
DyScO$_{3}$(110) at thermal equilibrium after 4 hours of annealing.
The lines in Fig.~\ref{fig:2}(a) represent the step edges, detected by the 
Canny filter. 

The step meandering of perovskite oxide step edges can be represented by a 
triangular undulation of amplitude $A$ and wavelength $\lambda$ as 
schematically shown in Fig.~\ref{fig:2}(b). The continuous wavy line labeled by 
$w$ is representing the step meandering of the step edge, corresponding to the 
white lines in Fig.~\ref{fig:2}(a) representing the DyScO$_3$(110) step edges 
in this example. The mean line of the meandering of the 
step edge is represented by $m$.

Now, by measuring the deviation-deviation correlation 
function $G(r)$ of the step one can extract the mean square kink length 
\cite{zandvliet2000}:
\begin{equation}
	G(r)=<\left(h(r)-h(0)\right)^2>=<k^{2}>\frac{r}{a_{0}}
\label{EqAFM:Gr}
\end{equation}
where $h(r)$ is the deviation measured in the direction perpendicular to the 
step edge at position $r$, $h(0)$ = $<h(r)>$ = 0 is the deviation measured 
in the direction perpendicular to the step edge at position 0, $r$ is the 
position along the high symmetry direction parallel to the step edge, 
and $a_{0}$ the unit cell length parallel to the step edge. Now $<k^{2}>$ is 
the mean-square kink length, see Fig.~\ref{fig:2}(b). 
From the mean square kink length, the kink energy can be calculated as:
\begin{equation}
 	<k^2>=
\frac{{\sum\limits_{k=-\infty}^\infty} 
k^2e^\frac{-E_{kink}(n)}{k_BT}}
{\sum\limits_{k=-\infty}^\infty e^\frac{-E_{kink}(n)}{k_BT}} 
\label{EqAFM:K2}
\end{equation}
where $k$ is the kink length, $k_{B}$ the Boltzmann constant and $T$ the sample 
temperature. The kink energy $E_{kink}$ is related to the nearest neighbour 
energy $E_{n}$ as: 
\begin{equation}
E_{kink}(n)=n \cdot E_n/2
\label{EqAFM:Ekinkk}
\end{equation}
By substitution, the mean square kink length $<k^{2}>$ and kink energy 
$E_{kink}$ relation can be simplified to:
\begin{equation}
 <k^{2}>=\frac{2y}{(1-y)^2}	
\label{EqAFM:KandY}
\end{equation}
where:
\begin{equation}
y=e^{\frac{-E_{kink}}{k_BT}}=e^{\frac{-E_n}{2k_BT}}
\label{EqAFM:EkinkY}
\end{equation}

Eqs. \ref{EqAFM:Gr}-\ref{EqAFM:EkinkY} reveal that the kink energy $E_{kink}$ 
can be extracted by measuring the slope $<k^{2}>$ of the correlation 
function for small enough $r$. The linear dependence in this range of the 
correlation function implies a random kink formation distribution.

In order to determine the strain energy, we fit the correlation function of 
the perovskite oxide step waviness by a triangular step undulation model. Note 
that the triangular step undulation is indicative of a strong energy relaxation 
mechanism along the step. There is a competition between two energy terms in 
minimizing the surface free energy, i.e. the energy cost to create additional 
step length versus the energy reduction by strain relaxation due to the 
triangular undulations along the step edge. 

The total free energy of a simple square-wave like step with periodicity 
$\lambda$ can be described as \cite{zandvliet2003}:
\begin{equation}
F=\gamma+\frac{2F_{step}}{\lambda}-\frac{(1-\upsilon)(\Delta\sigma)^{2}}{
\pi\mu\lambda}ln(\frac{\lambda}{2\pi a}\cdot sin(\pi p))
\end{equation}
where $\gamma$ is the free energy per unit area (in perovskite oxides, the free 
energy per unit area for a step-up and step-down domain are equal), $F_{step}$ 
is the step free energy per unit length, $\upsilon$ is the Poisson ratio, 
$\mu$ the bulk modulus 
and $\Delta\sigma$ the force monopole at the step edge. 

The total free energy per unit area of a triangular step edge can be calculated 
as
\begin{widetext}
\begin{equation}
F = \frac{1}{A}[\int\limits_{h=0}^A \gamma dh+\int\limits_{h=0}^A 
\frac{2F_{step}}{\lambda} dh 
- \int\limits_{h=0}^A\frac{2C}{\lambda}ln(\frac{\lambda}{2\pi a}) dh 
- \int\limits_{h=0}^A \frac{2C}{\lambda}ln(sin(\pi p))dh
\label{EqAFM:triangularwave}
\end{equation}
\end{widetext}
with the strain relaxation energy constant $C$ as,
\begin{equation}
C=\frac{(1-v)(\Delta\sigma)^{2}}{2\pi\mu}
\end{equation}
and,
\begin{equation}
p=\frac{A-h}{A}
\end{equation}
with $A$ the amplitude as shown in Fig.~\ref{fig:2}(b).

Now, Eq. \ref{EqAFM:triangularwave} can be rewritten as:
\begin{equation}
F=\gamma + \frac{2F_{step}}{\lambda} - \frac{2C}{\lambda}ln(\frac{\lambda}{2\pi 
a}) + \frac{2C}{\lambda\pi}\int\limits_{\pi}^0 ln[sin(x)]dx
\label{EqAFM:integrationfunction}
\end{equation}
The integral of Eq. \ref{EqAFM:integrationfunction} can be solved by using 
trigonometric identities and integration by substitution, such that
\begin{widetext}
\begin{eqnarray}\nonumber
\int\limits_{\pi}^0 ln[sin(x)]dx &=& \int\limits_{\pi}^0 
ln[2sin(\frac{x}{2})cos(\frac{x}{2})]dx \\ \nonumber
&=&  \int\limits_{\pi}^0 ln(2)dx + \int\limits_{\pi}^0 ln[sin(\frac{x}{2})]dx + 
\int\limits_{\pi}^0 ln[cos(\frac{x}{2})]dx \\ \nonumber	
&=& -\pi ln(2) + 2 \int\limits_{\frac{\pi}{2}}^0 ln[sin(y)]dy + 
\int\limits_{\pi}^0 ln[sin(\frac{x+\pi}{2})]dx	\\	\nonumber
&=& -\pi ln(2) + 2 \int\limits_{\frac{\pi}{2}}^0 ln[sin(y)]dy + 2 
\int\limits_{\pi}^\frac{\pi}{2} ln[sin(y)]dy	\\	
&=& -\pi ln(2) + 2 \int\limits_{\pi}^0 ln[sin(y)]dy 	
\label{EqAFM:derivationintegral}
\end{eqnarray}
\end{widetext}
Since $\int\limits_{\pi}^0 ln[sin(x)]dx=\pi ln(2)$, see Eq. 
\ref{EqAFM:derivationintegral}, the total free energy per unit area F of a 
triangular step edge is equal to:
\begin{equation}
F=\gamma+\frac{2F_{step}}{\lambda}-\frac{2C}{\lambda}ln(\frac{\lambda}{2\pi a}) 
+\frac{2Cln(2)}{\lambda}
\end{equation}
The critical periodicity $\lambda_{c}$ follows from the minimum free energy per 
unit area by calculating $\frac{dF}{d\lambda}$=0 as: 
\begin{equation}
\lambda_{c}= 2\pi a e^{(\frac{F_{step}}{C}+1+ln(2))}
\label{EqAFM:criticallambda}
\end{equation}

Besides the ln(2) term in the exponent, Eq. \ref{EqAFM:criticallambda} is 
identical to the relation derived by Alerhand \emph{et al.} 
\cite{alerhand1988}.

Now, by determining the critical periodicity $\lambda_c$ from the 
experimentally obtained topographic AFM images, together with the obtained 
$E_{kink}$ (extracted from the slope $<k^{2}>$ of the correlation function for 
small enough $r$), one can calculate the strain relaxation energy constant $C$.

\section{Energetics of SrTiO$_3$(001) and DyScO$_3$(110) surfaces}
\begin{figure}
\centering\includegraphics[width=0.8\linewidth]{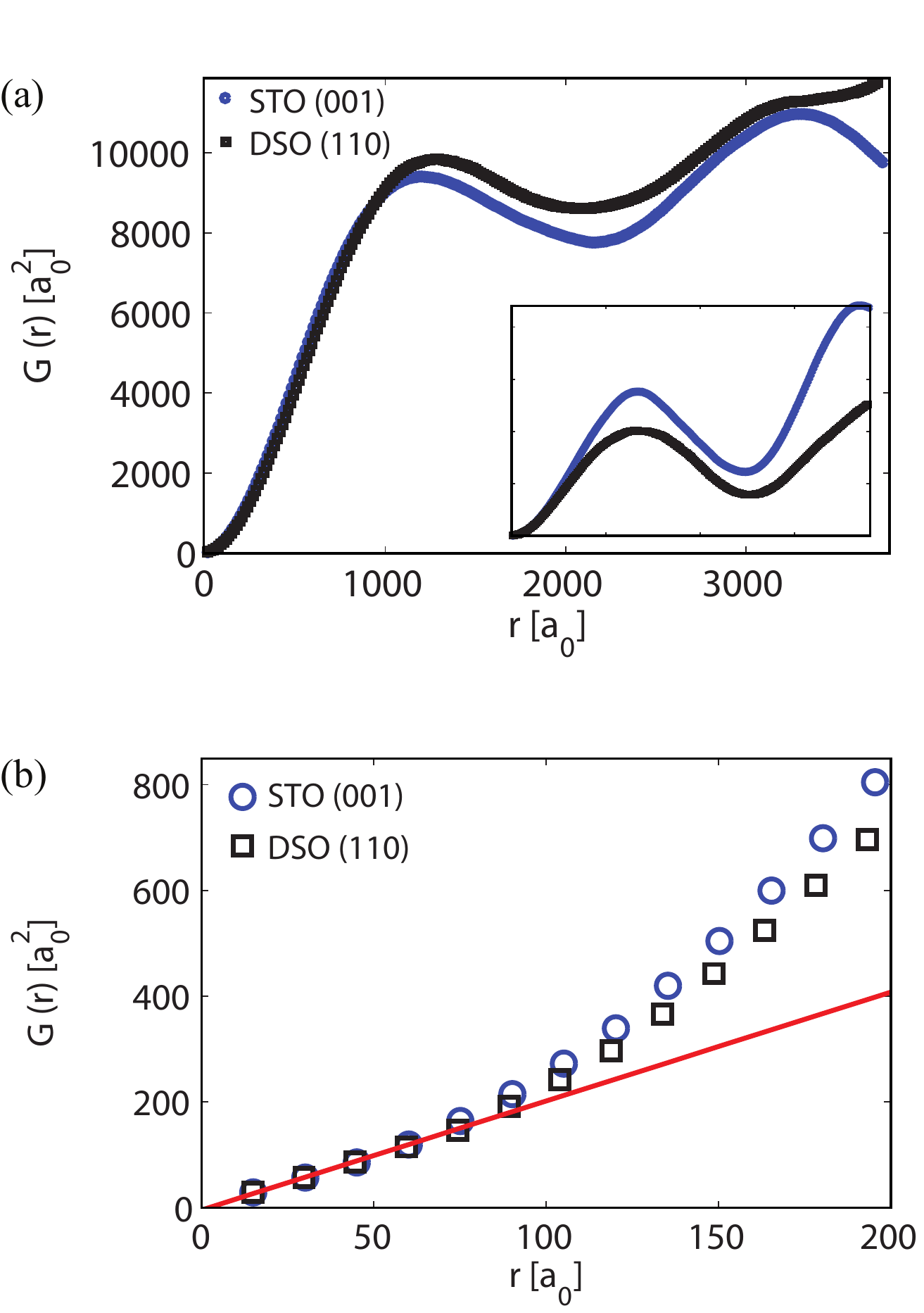}
\caption{Average correlation functions of SrTiO$_{3}$(001) (STO) and 
DyScO$_{3}$(110) (DSO) surface step edges (a). The inset shows a correlation 
function of a single step edge. 
The correlation functions show an average periodicity of $\lambda$ 
$\approx$~2100~$a_{0}$ $\approx$~830~nm for both surfaces, with only small 
variations from step edge to step edge. (b) The same correlation function for 
small $r$ ($r$=0~-~200~$a_{0}$). The average terrace widths of the samples 
investigated are of $<L>$ = 250~-~300~nm.}
\label{fig:3}
\end{figure}

In order to demonstrate how the energetics of vicinal perovskite surfaces can 
be studied by measuring its surface topography, we determine the $E_{kink}$ and 
strain relaxation for two prototypical perovskites i.e. SrTiO$_3$(001) and 
DyScO$_3$(110). For this, we performed the image analysis as described on 
Fig.~\ref{fig:1}(e) and (c) respectively, identifying the step edges by use of 
the described Canny filter. From these, we calculated the corresponding 
(average) correlation functions as depicted in Fig.~\ref{fig:3} which show 
surprisingly both qualitative and quantitave similarities for SrTiO$_3$(001) and 
DyScO$_3$(110). 
In Fig. \ref{fig:3}(a), the average correlation functions are shown for 
SrTiO$_{3}$(001) and DyScO$_{3}$(110) step edges with in the inset the 
correlation function of a single step edge. At large length scales, see Fig. 
\ref{fig:3}(a), the 
correlation functions of both systems show a clear correlation with an average 
periodicity $\lambda$, which 
slightly varies from step to step. Multiple minima and maxima are present 
demonstrating the strong correlation along the step edges. The ratio 
between minima and maxima amplitude $A$, see Fig.~\ref{fig:1}(b), slightly 
varies from step to step, however, the periodicity remains constant for both 
material systems. 

In Fig.~\ref{fig:3}(b), the correlation functions are plotted for small $r$ 
ranging from 0 - 200 $a_{0}$.
At short length scales, see Fig.~\ref{fig:3}(b), the correlation 
function starts to deviate from the linear fit ($G(r)$ $\sim$ $r$) around 
$r$~$>$~60~$a_{0}$ to a quadratic relation ($G(r)$~$\sim$~$r^{2}$).

From the determined correlation function, the critical periodicity 
$\lambda$$_{c}$ for both surfaces is found to be equal to 2100~$a_{0}$ 
(corresponding to $\approx$~830~nm).
Applying this to Eq. \ref{EqAFM:criticallambda} results in a ratio of 
$\frac{F_{step}}{C}$ = $\frac{E_{kink}}{C}$ = 4.13. 
Now, the kink formation energy can be determined  
from Fig.~\ref{fig:3}(b), a $<k^{2}>$=2.1$\pm$0.1$_{0}^{2}$ is extracted 
assuming random kink formation on this short length scale, which results in 
$E_{kink}$ = 0.10 eV/$a_{0}$ using Eqs. \ref{EqAFM:KandY} and 
\ref{EqAFM:EkinkY}.
Note, that from this the nearest neighbour energy $E_{n}$ can be determined 
as $E_{kink}$ = 0.21 $\pm$ 0.01 eV for SrTiO$_{3}$(001) and 
DyScO$_{3}$(110), both assumed as a (pseudo-)cubic crystal structure 
\cite{zandvliet2005}. Furthermore, the strain relaxation energy constant C 
can then be determined as 24 meV/$a_{0}$ for both 
surfaces. 
This value is close to the reported value range of
$E_{n}$ = 0.25 - 0.6 eV, commonly used in Kinetic Monte Carlo 
simulations to simulate thin film growth \cite{puiman2002}. 

\begin{figure}
\centering\includegraphics[width=\linewidth]{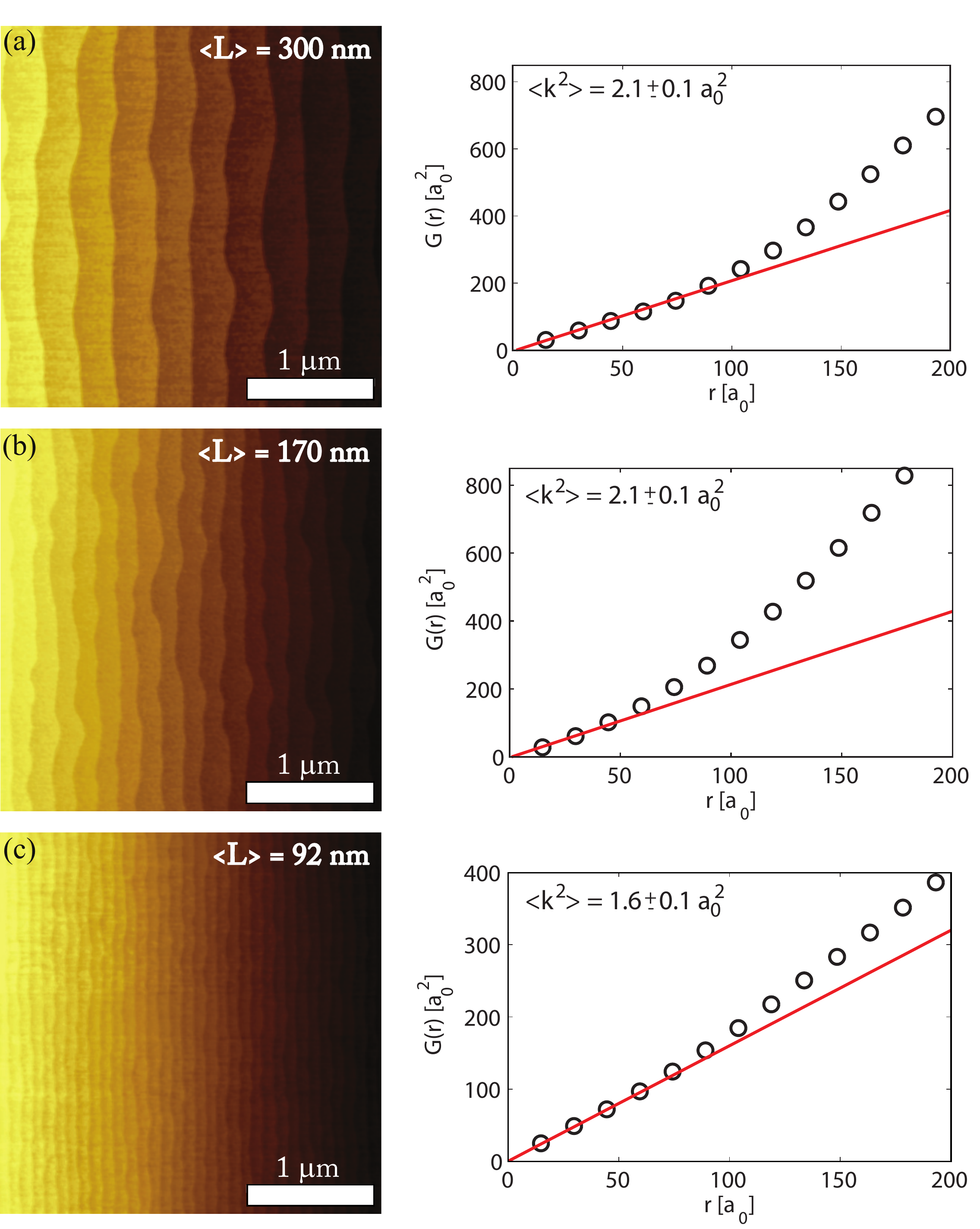}
\caption{\emph{Ex-situ} TM-AFM images of annealed DyScO$_{3}$(110) with 
different vicinal angles in the left panel labeled with their average terrace 
width. In the right panel the accompanying average correlation functions are 
plot with indicated mean square kink length.}
\label{fig:0}
\end{figure}

As mentioned before, the correlation along the surface steps 
depends on the average terrace width $<L>$. When the average 
terrace width becomes too small, the correlation function becomes 
less pronounced resulting in overestimating $E_{kink}$.
A small terrace width distribution as demonstrated for 
DyScO$_{3}$(110) is indicative of step-step interactions. However, in the case 
that step-step interactions are present, the step-edge morphology, 
correlation function and extracted energetic values can be influenced when 
terrace widths become so small that step edges start interacting. 
In Fig.~\ref{fig:0} we demonstrate this, by studying DyScO$_3$(110) samples 
with increasing vicinal cut. For an average terrace width of only 
$<L>$=92~nm, the step edge morphology is influenced, resulting in a skewed 
correlation function which in its turn results to an 
overestimated nearest neighbour energy.

For larger terrace widths of 300~nm and 170~nm, see 
respectively Fig.~\ref{fig:0}(a) and (b), a similar correlation function 
behavior is found on small length scales (0 - 100 $a_{0}$). At larger length 
scales (1000 - 3000 $a_{0}$) the correlation has less pronounced minima and 
maxima then for the DyScO$_{3}$(110) surface having smaller terrace widths 
($<L>$ = 
170~nm). 
The mean square kink length decreases with larger 
step-step repulsion as the entropy of steps decreases for small step-step 
separations. Typically, the influence of the step-step interaction potential on 
a 
wandering step increases with 1/$L^{2}$ \cite{zandvliet2000}. The decrease of 
the mean square kink length $<k^{2}>$ = 2.1 $\pm$ 0.1 $a_{0}^{2}$ towards lower 
values (we described $<k^{2}>$ = 1.6 $\pm$ 0.1 $a_{0}^{2}$ for $<L>$=92~nm 
here) results in a change of the found nearest neighbour energy of $E_{n}$ = 
0.21 $\pm$ 0.01 eV towards 0.23 $\pm$ 0.01 eV. 

The proposed model explains the quasi 1D character of a perovskite oxide step 
edge for terrace widths larger than 170 nm. Smaller terrace widths show a more 
2D character caused by the step-step interaction, which requires a more 
sophisticated model. Both, the entropic step-step interaction \cite{gruber1967} 
and step interaction energy models \cite{alerhand1988} used to describe the 
interaction of Si step-edges are not valid for DyScO$_{3}$(110) step-edges. The 
models assume a meandering Si step-edge catched between two straight 
step-edges, while the triangular undulation of step-edges on a DyScO$_{3}$(110) 
surface behaves coherent.

\section{Conclusions}
In conclusion, we demonstrated a method to determine both, the strain 
relaxation energy together with the step edge formation energy on a 
perovskite surface by measuring its vicinal surface topography by use of AFM.
Remarkably, we find similar triangular step edge undulations for two different 
perovskite surfaces at thermal equilibrium. From these we determine an average 
step edge undulation periodicity of $\approx$~2100~$a_{0}$ for the two 
different perovskite materials i.e. SrTiO$_{3}$(001) and DyScO$_{3}$(110). These 
step undulations are caused by strain relaxation along the step direction, 
determined to be 24meV/$a_0$. 
From the slope of the correlation function we determine also similar kink 
formation energies for both perovskite materials to be 0.10~eV/$a_0$ and 
corresponding nearest neighbour energy of 0.21$\pm$0.01~eV, in good agreement 
with values used in Kinetic Monte Carlo simulations of thin film growth 
\cite{puiman2002}.

From the similar step meandering along the step edges of the perovskite oxides 
studied and described here, we suggest that the surface energetics might depend 
on the local oxygen coordination. According to this, the surface energetics 
should change with different alternating planes / oxygen coordination, for 
example in the case of SrTiO$_{3}$(001)/AO - BO$_{2}$, SrTiO$_{3}$(110)/ABO - 
O$_{2}$ and SrTiO$_{3}$(111)/AO$_{3}$ - B planes, which is supported by DFT 
calculations and difference in required annealing times \cite{woo2015}. The 
importance of the oxygen coordination is also supported by the specific 
terminating plane at the surface, since it has been shown that nanostructures 
can be grown selective on mixed terminated substrates 
\cite{kuiper2011,sanchez2014}.

\begin{acknowledgments}
This work is part of the research programme of NanoNext NL project 9A 
nanoinspection and characterization, 
project 07 real-time atomic force microscopy growth monitoring during pulsed 
laser deposition of oxides.
This work is part of the research programme of the Foundation for Fundamental 
Research on Matter (FOM), which is financially supported by the Netherlands 
Organization for Scientific Research (NWO).
\end{acknowledgments}

\bibliography{./references.bib}

\end{document}